\documentclass[conference]{IEEEtran}
\usepackage{amssymb}
\usepackage{amsfonts}
\usepackage{graphicx}
\usepackage{array}
\usepackage{cite}
\usepackage{psfrag}
\usepackage{stfloats}
\usepackage{amsmath}
\usepackage{url}
\usepackage{times}
\usepackage{subfigure}

\begin{document}

\title{An Efficient Network Coding based Retransmission Algorithm for Wireless Multicast}
\author{Jalaluddin Qureshi, Chuan Heng Foh and Jianfei Cai \\
School of Computer Engineering\\
Nanyang Technological University, Singapore\\
\{jala0001, aschfoh, asjfcai\}@ntu.edu.sg}

\maketitle

\begin{abstract}
Retransmission based on packet acknowledgement (ACK/NAK) is a
fundamental error control technique employed in IEEE 802.11-2007
unicast network. However the 802.11-2007 standard falls short of
proposing a reliable MAC-level recovery protocol for multicast
frames. In this paper we propose a latency and bandwidth efficient
coding algorithm based on the principles of network coding for
retransmitting lost packets in a single-hop wireless multicast
network and demonstrate its effectiveness over previously proposed
network coding based retransmission algorithms.
\end{abstract}

\section{Introduction}
\label{sect:Introduction}One-to-many (broadcast/multicast)
transmission scheme is popular for many applications, and is widely
implemented in Wireless Local Area Networks (WLANs) for its
effectiveness in bandwidth consumption in a spectrum-limited
wireless space. WLANs transmission is currently dictated by
standards set out by IEEE 802.11-2007~\cite{ieee}. For one-to-one
(unicast) wireless transmission, transmission reliability is
achieved through Automatic Repeat Request (ARQ) variants or/and
Forward Error Correction (FEC) schemes. Since broadcast is a special
case of multicast, without loss of generality we will use the term
multicast henceforth. However for a multicast, no consideration is
made for ACK/NAK and RTS/CTS packet exchange in 802.11-2007 except
for those frames sent with the To DS field set. Additionally, for
multicast network where consideration for control packet is made,
such packets are collected individually one-by-one, and so is the
retransmission of the lost packets done, that is, one-by-one. As
such for multicast network, the reliability problem is two-folded:
1) Efficient mechanism for the transmission of control packets
(ACK/NAK, RTS/CTS), and 2) efficient retransmission of packets lost.

As multicasting is gaining popularity for applications such as file
distribution and multimedia conferencing, a more reliable scheme is
needed for the fulfillment of future growth in multicast network.
Motivated by promising applications of Network Coding (NC), recent
works\cite{conext-start} -\cite{iccsc-end} have demonstrated the
suitability of NC for retransmission of lost packets to improve
bandwidth performance in a multicast network.

Our algorithm is based on the concept of network
coding~\cite{nc},~\cite{cope}. Network coding in its simplest form
exploits the fact that rather than transmitting wireless packets
individually to some receivers which may be `overheard' by some
other receivers already having those packets, and vice versa, it is
often possible to combine those packets using bit-by-bit XOR
(denoted by $\oplus$) and transmit it as a single coded packet,
which can then be decoded by all (/most) of the receivers based on
the packets they already have. For illustration consider that
receiver $R_1$ has packet $c_1$ but not $c_2$, while $R_2$ has $c_2$
but not $c_1$. Rather than transmitting these two packets
individually, the transmitter can encode $c_1$ and $c_2$ to generate
$c_1\oplus c_2$, which is then multicast to both the receivers and
decoded.

The remaining paper is organized as follow: In
Section~\ref{sect:relatedwork} we give an overview of related work,
followed by the problem statement in Section~\ref{sect:problem}.
Following that, we discuss previously proposed coding algorithm in
Section~\ref{sect:coding} and our BENEFIT algorithm in
Section~\ref{sect:benefit}. We then confirm the performance of
BENEFIT with simulation results in Section~\ref{sect:simulation},
and finally present conclusion in Section~\ref{sect:conclusion}.

\section{Related Work}\label{sect:relatedwork}
Packet retransmission based on network coding for a one-to-many,
single-hop multicast network is a recent field of study, first
proposed by D. Nguyen et al.~\cite{netcod}, which was later further
elaborated into~\cite{vt} by D. Nguyen et al. In~\cite{vt} the
authors demonstrate bandwidth effectiveness achieved by employing
greedy network coding for retransmission over traditional ARQ
schemes through simulation work. In~\cite{conext-start} the authors
follow up the work in~\cite{netcod} by comparing various packet
coding algorithms for packet retransmissions. While in~\cite{majid},
the authors presents an analytical work on the reliability
performance of network coding compared with ARQ and FEC in a lossy
network. Network Coded Piggy Back (NCPB)~\cite{ncpb} demonstrates an
efficient and practical testbed implemented random linear network
coding based many-to-many reliable network model for real-time
multi-player game network.

Since our work primarily focuses on proposing an efficient network
coding based retransmission algorithm for a one-to-many single-hop
network, we will be comparing our results with the algorithm given
in~\cite{conext-start} which is the most closely related work.

\subsection{Our Contribution}
The novelty of our work is the development of a computationally
feasible network coding based retransmission algorithm whose gains
are two-folded: 1) Our algorithm BENEFIT delivers better throughput
with respect to the current best single-hop, NC based retransmission
algorithm, and 2) we also demonstrate that our algorithm achieves
minimum time to decode packets. None of the previous works
\cite{conext-start} -\cite{iccsc-end} on NC based retransmission
incorporates consideration of packet latency in their work.

Here we will also show that it is no longer necessary to follow the
packet coding rule~\cite{cope},~\cite{ncrouting} strictly. This
relaxation in the coding rule has the potential for modification and
development of other network coding based applications.

\section{Problem Statement}\label{sect:problem} Consider a
single-hop multicast network with $M$ fixed receiver stations $R_i$,
($i$ is the receiver station ID, $1\leq i\leq M$) with static
membership and $M\geq 2$, and a single transmitting station $T_x$.
Packet batch size is denoted by $N$. Packet reception at $R_i$
follows Bernoulli model, whereby a successful reception of packet
$c_k$ ($k$ is the datagram packet ID, $1\leq k\leq N$) at $R_i$ is
indicated by `0' and packet loss by `1' in the transmission matrix
(see Table~\ref{table:matrix}). A transmission matrix is a
2-dimensional array table, where the rows represents $R_i$ and
columns represents $c_k$. For a given packet, its packet utility
$cu_k$ ($0\leq cu_k\leq M$) is defined as the number of receiver(s)
which have not received the packet (i.e. the numbers of `1's in a
given column). While the receiver utility $ru_i$ indicates the
number of packet(s) not received by the receiver $R_i$ from a given
set of specified packet(s). For Bernoulli model, packet loss at all
receivers is homogeneous, and is determined by a fixed loss
probability $p_i$, which gives a packet successful reception
probability of $1-p_i$. For a fixed batch size, $L_i$ denotes the
number of lost packets for station $i$. $Q_j$ is the probability
that after $N$ transmissions, the total number of packets lost is no
more than $j$ ($1\leq j\leq N$).

The time taken for one transmission is represented by one time slot.
The \emph{time to decode} a lost packet for a given $R_i$ is the
total number of transmissions (original transmission, retransmission
and transmission of coded packet) after which the lost packet is
recovered by the given $R_i$.

For simplicity we assume that there is a reliable control packet
exchange mechanism in the network\footnote{Control packets in
multicast network can be implemented by designing ACK packets from
multiple STA such that, upon reception of these simultaneously
transmitted ACK packets, the original sender is able to efficiently
decode the packet which is the superimposition of all ACK packets
and infer which receiver STA have received the datagram
packet~\cite{epfl}.} and that all coded/retransmitted packets are
successfully received by the receivers. In the context of BENEFIT
algorithm, a \emph{benefit} value is generally defined as the number
of `1's in the transmission matrix which are converted to `0's after
the transmission of a coded packet or retransmission of the packet,
hence the name of the algorithm: `BENEFIT'.

\subsection{Theoretical Numbers of Retransmissions}
The probability that $L_i\leq j$, for a single $R_i$ is given by
\begin{equation}
P[L_i\leq j] = \sum_{c=0}^j (_{c}^N) p_{i}^c (1-p_i)^{N-c}.
\end{equation}

The probability that all $M$ stations experience a packet loss rate
no more than $j$ is given by $\prod_{i=1}^M P[L_i\leq j]$. Given
this result, the probability that the total number retransmission is
$j$, is given by

\begin{equation}
Q_j = \prod_{i=1}^M P[L_i\leq j] - \prod_{i=1}^M P[L_i\leq j-1].
\end{equation}

A more elaborative discussion of retransmission bandwidth for
different transmission schemes compared with network coding is given
in~\cite{vt},~\cite{majid}.

\section{Coding Algorithm}\label{sect:coding}
Previous coding algorithms (except random linear network coding,
RLNC) were build on the foundation of a simple \emph{packet coding
rule}~\cite{cope},~\cite{ncrouting}:

\emph{ For $T_x$ to transmit (/retransmit) $M$ packets $c_1$, ...,
$c_M$ to $M$ receivers, $R_1$, ..., $R_M$ respectively, the coded
packet obtained by coding $M$ packets $c_1$, ..., $c_M$ can only be
decoded at $R_i$ if $R_i$ has ($M-1$) of $c_j$ packets, except $c_i$
($j\neq i$).}

We now discuss the major coding algorithm used in network coding
literature.

\subsection{Greedy Network Coding} A greedy algorithm for coding
packets has been traditionally used in several network coding based
literature and still continues to be a dominant approach in many NC
networks like IP-level routers in the Internet~\cite{nc}, wireless
mesh network~\cite{cope} and multi-hop wireless
routing~\cite{ncrouting}. A greedy network coding algorithm, like
traditional greedy algorithm makes coding decisions which gives
optimal local results. A greedy coding algorithm encodes current
locally available packets iteratively as long as it can be decoded
by all the intended receivers, without consideration whether its a
`globally' optimal solution or not.

\subsection{Random Linear Network Coding}
RLNC~\cite{ncpb} is a decentralized network coding approach, whereby
the coded packet is given by $c_{coded}$ = $\sum_{k=1}^{k=N}$ $g(e)
c_k$, where $g(e)$ is the global encoding vector, and is included in
$c_{coded}$ as an overhead information in the packet header. Each of
the receiver $R_i$ must successfully receive $N$ innovative packets
(i.e. coded packets which are linearly independent of the previously
received coded packet). Once the receivers have $N$ innovative
packet, it can then decode $N$ packets using simple matrix
inversion.

\subsection{Sort-by-Utility}
The coding algorithm which E. Rozner et. al.~\cite{conext-start}
proclaims to be the delivering the best performance for a one-hop
multicast network is Sort-by-Utility. Therefore we will be comparing
our BENEFIT algorithm with Sort-by-Utility for evaluation purposes.
In a Sort-by-Utility coding algorithm, the $T_x$ first transmits $N$
packets, and then sorts the packets in descending order of their
$cu_k$ values, using arrival time as tie-breaker for those packets
have equal packet utility. Once the packets are sorted, the
remaining operation of Sort-by-Utility is essentially a greedy
coding algorithm, i.e. the $T_x$ then iteratively starts coding
successive packets starting from packets having highest packet
utilities and codes them with successive sorted packets as long as
the coded packet can be decoded by \emph{all} receivers.

\begin{table}[htbp]
\caption{Transmission Matrix Example} \label{table:matrix}
\begin{center}
\begin{tabular}{|c|c|c|c|c|c|c|}
\hline $R_i$/$c_k$ & $ru_i$& $c_1$ & $c_2$ & $c_3$ & $c_4$ & $c_5$ \\
\hline $cu_k$      &   10  &   2   &   3   &   1   &   2   &   2   \\
\hline $R_1$       &   3   &   1   &   1   &   0   &   0   &   1   \\
\hline $R_2$       &   2   &   0   &   1   &   0   &   1   &   0   \\
\hline $R_3$       &   2   &   0   &   1   &   1   &   0   &   0   \\
\hline $R_4$       &   3   &   1   &   0   &   0   &   1   &   1   \\
\hline
  \end{tabular}
\end{center}
\end{table}

Consider as an example the matrix given in Table~\ref{table:matrix}.
Sort-by-Utility algorithm after initial packet transmission ($c_1$
to $c_5$) would sort the transmitted packets based on its packet
utilities $cu_k$, i.e. $c_2$, $c_1$, $c_4$, $c_5$, $c_3$. Then $T_x$
transmits the packets as follows: $c_2$, $c_1\oplus c_3$, $c_4$,
$c_5$. Thus requiring a total of 4 retransmissions with an average
\emph{time to decode} a packet to be 4.4 time slots.

\section{BENEFIT Algorithm}\label{sect:benefit}

\begin{figure}
\begin{center}
\includegraphics[width = 0.5\textwidth]{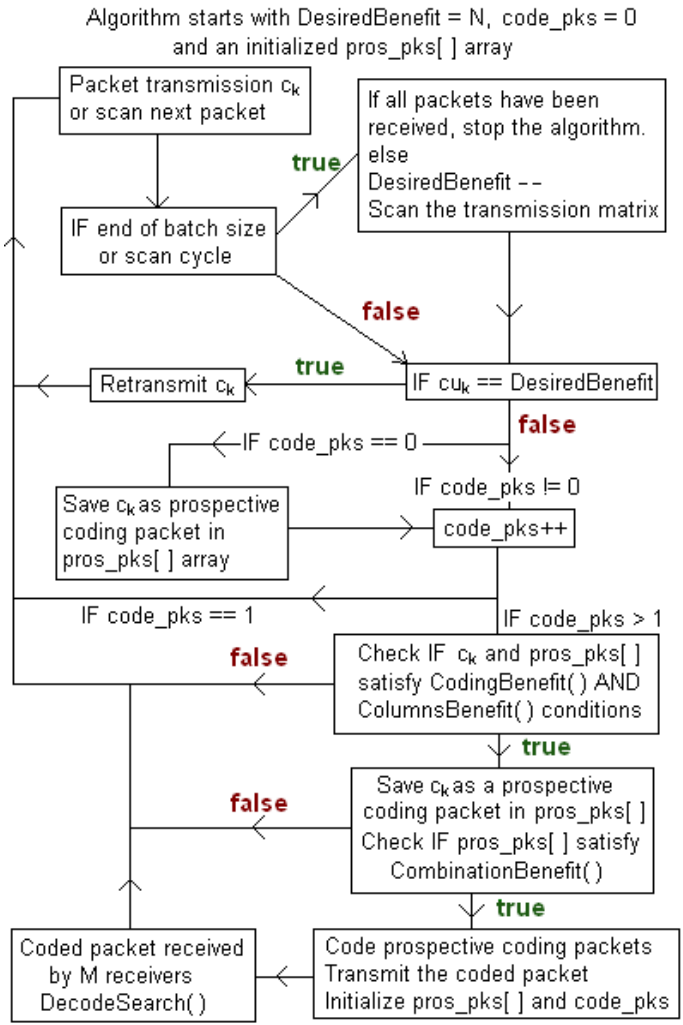}
\end{center}
\caption{BENEFIT algorithm - flowchart} \label{flowchart}
\end{figure}

BENEFIT (Fig~\ref{flowchart}), unlike Sort-by-Utility does not need
to wait until the end of the batch size ($N$ packet transmissions)
before starting the retransmission process. It start transmitting
coded packet once the prospective coding packets satisfy the
following three conditions: $CodingBenefit()$, $ColumnsBenefit()$
and $CombinationBenefit()$ (see Table~\ref{table:terms}).
Retransmitting as soon as the right conditions are met rather than
wait till the end of the batch size in effect reduces the time to
decode the packet. BENEFIT works on the basis that \emph{it is not
necessary for the coded packet to be decodable by all the receivers
immediately}, assuming that the non-decodable coded packet can be
decoded based on future transmission of coded packet(s). This
principle is in essence the key strength of BENEFIT and thus, this
way it contrasts the traditional \emph{packet coding rule}.

For the first scan cycle, to decide whether to transmit a packet
$c_k$ or to scan the next packet (see the first step of
Fig.~\ref{flowchart}) is decided based on the fact that if any
previously transmitted packet has never been the first prospective
coding packet (i.e. pros\_pks[0]) in that cycle, and the
\emph{current} value of $cu_k$ of that previously transmitted packet
is $1\leq cu_k <M$, then the algorithm scan the next packet and
stores it as the first prospective coding packet, else it transmits
the next packet. For consecutive cycles, the algorithm only scans
the packets. $CodingBenefit()$ ensures that packets are only coded,
if the immediate benefit derived from such coding outweighs or
equals the benefit derived from transmitting a single packet
(uncoded) with minimum packet utility ($cu_k$) from the set of
prospective coding packets and considered packet ($c_k$).
$ColumnsBenefit()$ selects the most suitable (/fittest) packets for
coding, by eliminating those packets which can not be decoded by at
least one receiver STA immediately. If the packets satisfy
$CodingBenefit()$ and $ColumnsBenefit()$ conditions but not
$CombinationBenefit()$, then they are considered eligible
prospective coding packets for combination with other packet(s), and
thus the algorithm then searches (scans) for other packet(s), which
in combination with the previous prospective coding packets will
satisfy the three conditions for packet coding. If the algorithm
reaches the end of the batch size and there are still `1's in the
transmission matrix, then the $CombinationBenefit()$ condition is
relaxed by decrementing $DesiredBenefit$ and the algorithm then
starts a new scan cycle (maximum of $M$-1 scan cycles) until the
transmission matrix is composed of $M$*$N$ `0's.

\begin{table}
\caption{Definition of BENEFIT terms} \label{table:terms}
\begin{tabular}{|l|}

\hline
$CodingBenefit()$ \\

Checks the following equality: \\
$DecodeBenefit()\geq$ $MinimumBenefit()$ \\

\hline
$DecodeBenefit()$ \\

Finds out how many receivers STA will benefit \emph{immediately} from the \\
transmission of the coded packet, $0\leq DecodeBenefit()\leq M$.\\

\hline
$MinimumBenefit()$ \\

From the list of prospective coding packets, find $c_k$ with minimum\\
current packet utility $cu_k$, $0\leq MinimumBenefit()\leq M$.\\

\hline
$ColumnsBenefit()$\\

From the list of prospective coding packets, checks if every packet can\\
be decoded by at least one receiver STA immediately.\\

\hline
$CombinationBenefit()$\\

From the list of considered packets for coding, calculates the number of\\
receiver STA which will benefit either \emph{immediately} ($ru_i$=1) or in \emph{future}\\
($ru_i\geq$ 2) from the coding of the prospective coding packets. And then\\
checks if its equal to $DesiredBenefit$ (see Fig~\ref{flowchart}). $ru_i$ value in the\\
context of $CombinationBenefit()$ is computed only for the packets\\
in pros\_pks[ ], $CombinationBenefit()\geq DecodeBenefit()$.\\

\hline

$DecodeSearch()$\\
Decodes the arrived coded packet if it can. If the packet gets decoded,\\
then search the memory for any previously non-decodable packet which\\
can now be decoded based on the current decoded packet, and decode it.\\

\hline
$Benefit$ $immediately$/ $Immediate$ $decoding$\\

The decoding of coded packet `on the spot,' without the need for\\
information from future transmission(s).\\

\hline
  \end{tabular}
\end{table}

\subsection{Computational Complexity of BENEFIT}
The computational complexity of $CodingBenefit()$,
$ColumnsBenefit()$ and $CombinationBenefit()$ all grow linearly with
respect to the number of prospective coding packets and receivers.
While $DecodeSearch()$ can be implemented using a binary search
algorithm whose average complexity is logarithmic. Given that the
number of prospective coding packets increases with the number of
receivers, the computational complexity of BENEFIT can be considered
to be linearly increasing with the number of receiver STAs.

\subsection{Illustrative example - BENEFIT}
Consider the algorithm given in Fig.~\ref{flowchart}, illustrated
with the transmission matrix given in Table~\ref{table:matrix}.
After the $T_x$ transmit $c_1$, $c_1$ is stored as the first
prospective coding packet. $T_x$ then transmit $c_2$, $c_1$ and
$c_2$ are then checked for $CodingBenefit()$ and $ColumnsBenefit()$
conditions, which they satisfy. Hence they are then checked for
$CombinationBenefit()$ condition. Since both $c_1$ and $c_2$ satisfy
the $CombinationBenefit()$ condition as well, the packets are coded
and transmitted. Only $R_1$ is not able to decode $c_1\oplus c_2$
immediately (current value of $cu_1$=$cu_2$=1). The algorithm then
scan the next packet $c_2$ and stores it as the first prospective
coding packet, and the $T_x$ then transmit $c_3$. Since $c_2$ and
$c_3$ satisfy $CodingBenefit()$ and $ColumnsBenefit()$ conditions
but not $CombinationBenefit()$ condition, $c_3$ is therefore saved
as a prospective coding packet. The $T_x$ then transmits $c_4$,
which in addition to $c_2$ and $c_3$ satisfy $CombinationBenefit()$
condition. Hence the packets are coded and transmitted. All
receivers are able to immediately benefit from the transmission of
$c_2\oplus c_3\oplus c_4$. $DecodeSearch()$ function at $R_1$ after
decoding $c_2\oplus c_3\oplus c_4$, decodes $c_1\oplus c_2$ using
$c_2$ and obtains $c_1$. The last packet in the batch $c_5$ is then
transmitted and stored as prospective coding packet, however as it
is obvious, there is not any possibility of finding coding packets
for $c_5$, the algorithm decrements the value of $DesiredBenefit$
twice, following which $c_5$ is retransmitted without any encoding
in the third scan cycle.

This example illustrates that BENEFIT requires a total of only 3
retransmissions ($c_1\oplus c_2$, $c_2\oplus c_3\oplus c_4$ and
$c_5$) in contrast to 4 retransmissions used by Sort-by-Utility (see
section~\ref{sect:coding}-C), with an average \emph{time to decode}
a packet of 1.9 time slots ($c_1\oplus c_2$ transmitted after the
transmission of $c_2$, and $c_2\oplus c_3\oplus c_4$ transmitted
after the transmission of $c_4$) in contrast to 4.4 time slots used
by Sort-by-Utility. Hence for this example, it has been shown, that
BENEFIT outperforms Sort-by-Utility both in terms of retransmission
bandwidth and packet delay.

\section{Simulation Results}\label{sect:simulation}

\begin{figure}
\begin{center}
\includegraphics[width = 0.5\textwidth]{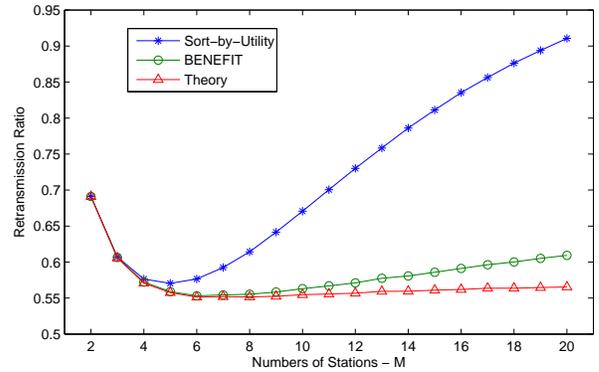}
\end{center}
\caption{Retransmission ratio against $M$, for $p_i$=0.5, and
$N$=200} \label{fig:1}
\end{figure}

\begin{figure}
\begin{center}
\includegraphics[width = 0.5\textwidth]{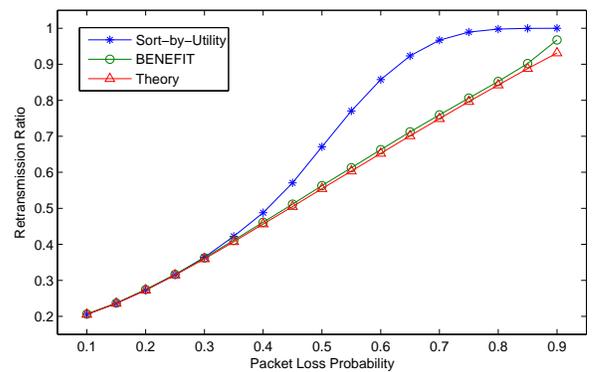}
\end{center}
\caption{Retransmission ratio against $p_i$, for $M$=10, and
$N$=200} \label{fig:2}
\end{figure}

We construct a C++ based discrete time simulator, using Random
Number Generator to generate transmission table like the one given
in Table~\ref{table:matrix}. The characteristics of the network
shall be the same as mentioned in Section~\ref{sect:problem}. For
each set of values, the simulation is repeated 1000 times. For
performance evaluation, we use \emph{retransmission ratio} (also
used in~\cite{conext-start}) which is defined as the total number of
retransmissions using coding algorithm divided by the total number
of retransmissions using traditional 802.11 retransmission scheme.
Theory in Fig.~\ref{fig:1} and~\ref{fig:2} refers to retransmission
ratio obtained by dividing $Q_j$ (derived in
Section~\ref{sect:problem}-A) with the total number of
retransmissions using traditional 802.11 retransmission scheme.

\begin{figure}
\begin{center}
\includegraphics[width = 0.5\textwidth]{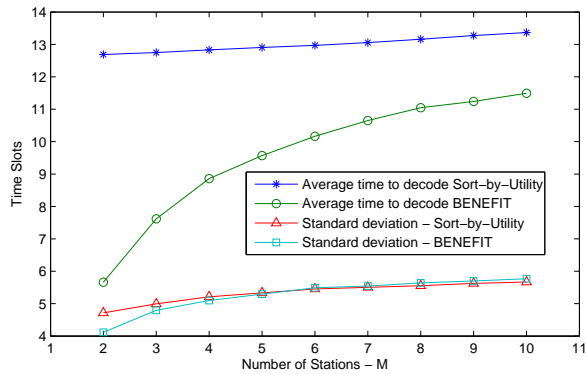}
\end{center}
\caption{Time slots against $M$, for $p_i$=0.25, and $N$=20}
\label{fig:3}
\end{figure}

Figure~\ref{fig:1} shows that BENEFIT consistently performs better
than Sort-by-Utility. The initial trough in the graph is because of
more coding opportunity available with an increase in number of STA.
A simple heuristic explanation for this is that for 2 STA, there is
scope for only 2 packets to be coded, however for 4 STA, there is
scope for 2, 3 and 4 packets to be coded together based on
opportunities. However as the number of STA increases further,
retransmission ratio starts increasing as then, increase in conflict
opportunities (as shown in Fig.~\ref{fig:3}, as the number of STA
increases, the time to search for prospective coding packet also
increases) between coding packets outweighs increase in coding
opportunities. Figure~\ref{fig:2} shows the performance of BENEFIT
over a range of loss probability values. Figure~\ref{fig:1}
and~\ref{fig:2} proves that for a small-medium network BENEFIT
performs close to the theoretical bound for all ranges of $p_i$.

\begin{figure}
\begin{center}
\includegraphics[width = 0.5\textwidth]{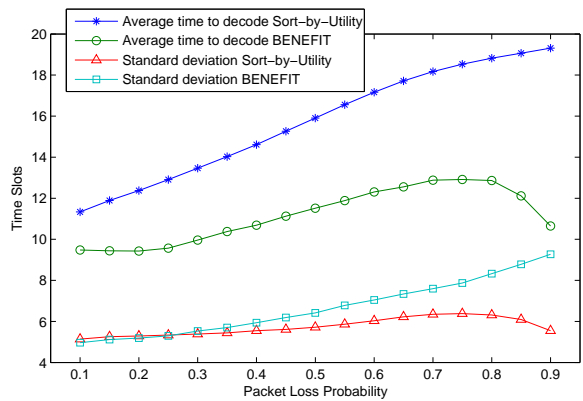}
\end{center}
\caption{Time slots against $p_i$, for $M$=5, and $N$=20}
\label{fig:4}
\end{figure}

While the bandwidth performance of BENEFIT and Sort-by-Utility is
almost similar for low loss probability and/or small network, for
such networks BENEFIT can still be useful for real-time applications
which are highly delay sensitive. As Fig.~\ref{fig:3} shows that
even for a small batch size, the average time BENEFIT takes to
decode/retransmit a packet is far less than that of Sort-by-Utility.
BENEFIT latency efficiency can be improved further by decrementing
the initial value of $DesiredBeneft$, which will relax the
$CombinationBenefit()$ condition and thus require the $T_x$ to spend
less time searching for suitable coding packet. Decreasing the batch
size also reduces packet retransmission delay as show
in~\cite{conext-start},~\cite{vt}. However both these techniques
will come at a tradeoff cost of an increase in retransmission ratio.
Flexibility to balance throughput-delay tradeoff in BENEFIT allows
the network designer to modify the algorithm based on the network
requirements. Figure~\ref{fig:4} shows that the average time to
decode packet gradually increases with $p_i$ for BENEFIT, however as
$p_i$ crosses 0.8 the average time to decode packet starts
decreasing as most of the packets satisfy $cu_k==DesiredBenefit$
condition and hence are retransmitted without any coding. The time
saved searching for prospective coding packets reduces the average
time to decode. This also explain an increase in standard deviation
for BENEFIT in Fig.~\ref{fig:4}, as some packets are retransmitted
without any encoding (shorter waiting time), while other packets
need to wait for longer to find suitable coding partners.

\section{Conclusion}\label{sect:conclusion}
In this paper we have demonstrated a computationally feasible,
bandwidth and latency efficient retransmission coding algorithm,
which doesn't strictly follows the traditional coding rule.
Selectively modifying the BENEFIT conditions would also allow the
network designer to adjust the algorithm as per the throughput-delay
requirements of the network. We believe that there is potential for
research work to exploit relaxation in the coding rule, and study
modifications to mechanisms like COPE~\cite{cope} based on rules
derived from BENEFIT.

\end{document}